\documentstyle[prl,aps,multicol]{revtex}
\input epsf

\begin{document}

\title{Methods for Reliable Teleportation}

\author{Lev Vaidman and Nadav Yoran} 

\address
{School of Physics and Astronomy, 
Raymond and Beverly Sackler Faculty of Exact Sciences, \\
Tel-Aviv University, Tel-Aviv 69978, Israel.} 

\date{}

\maketitle

\begin{abstract}
Recent experimental results and proposals towards  implementation
of quantum teleportation are discussed. It is proved that  reliable
(theoretically, 100\% probability of success)  teleportation cannot be achieved
using the methods applied in recent experiments, i.e., without quantum systems
interacting one with the other. 
Teleportation proposal involving atoms and electro-magnetic cavities 
are reviewed and the most feasible methods are described. In particular,
 the language of nonlocal measurements has been applied which has
 also been used for
presenting a method for teleportation of quantum
states of systems with continuous variables. 
\end{abstract}


\begin{multicols}{2}

\section{Introduction}
\label{Intro}

 Teleportation is ``... apparently instantaneous transportation
of persons etc., across space by advanced technological
means''\cite{OX}.  Recently, Bennett, Brassard, Crepeau, Jozsa, Peres,
and Wootters (BBCJPW) \cite{BBC} proposed a gedanken experiment which
they  called ``quantum teleportation''.
 Many proposals were suggested  and a few  real experiments
 \cite{Inn,Rome,Inn-mix}
 were performed inspired by the BBCJPW work.
 However, an experiment, demonstrating a reliable (i.e., 100\%
 successful, given ideal devices) teleportation of an unknown state of
 an external quantum system has not been performed yet.
 In this paper we  prove
that the methods used in the current experiments cannot lead to reliable
teleportation and   discuss other proposals for
teleportation.

Quantum teleportation transports the quantum state of a system and/or
its correlations to another system.  The state is disintegrated in one
place and a perfect replica appears at a distant site.  The state or its
complete description is never located between the two sites during the
transportation. The teleportation procedure, apart from   quantum
channels (prepared in
advanced), requires telegraphing surprisingly small
amount of information between the two sites. All these properties justify the
name ``teleportation'' for the BBCJPW procedure. Note that
telegraphing  classical information cannot be {\em instantaneous}.
Therefore, teleportation cannot be instantaneous too.  This, however, is not
surprising, since quantum states can carry information and  
special relativity does not allow instantaneous transmission of
signals.

The organization of this paper is as follows. Section \ref{BBCJPW} reviews the original teleportation method. Section
\ref{proof} is devoted to the proof that without interaction between
``quantum'' objects the  measurement of the  nondegenerate Bell operator, which is the core of the
original proposal, cannot be performed.   Section \ref{tele-int}
analyzes teleportation methods which use quantum-quantum interactions,
in particular, a method based on nonlocal measurements \cite{tele-V}.
In Section \ref{tele-exp} we discuss arguably a more promising
experimental proposals for reliable teleportation including
implementation of ``crossed'' nonlocal measurements scheme for
for teleportation of quantum states of atoms. In Section \ref{tele-cont} we discuss
proposals for teleportation of quantum states of continuous
variables. Section \ref{conc} presents a discussion which concludes
the paper.

\section{The original  BBCJPW teleportation procedure}
\label{BBCJPW}

Let us start with the analysis of teleportation of a quantum state of
a two-level system.  The states under discussion are: spin states of a
spin-$1\over 2$ particle, $|{\uparrow}\rangle$ and
$|{\downarrow}\rangle$, polarization states of a photon,
$|{\leftrightarrow}\rangle$ and $|{\updownarrow}\rangle$, the states
of a photon in a two-arm interferometer, $|a\rangle$ and $|b\rangle$,
ground and excited state of an atom (or ion), $|g\rangle$ and
$|e\rangle$, photon number states of a microwave cavity, $|0\rangle$
and $|1\rangle$.  Mathematically, there is no difference between the
analysis of various two-level systems. For the present analysis,
following the tradition, we will use spin states, in spite of the fact
that this is the only system from the list above for which there is no
proposal for a real experiment.

The original BBCJPW teleportation procedure consist of three main
stages:

i) Preparation of an EPR pair
\begin{equation}
|EPR\rangle = {1\over \sqrt2}(|{\uparrow}\rangle|{\downarrow}\rangle -
|{\downarrow}\rangle|{\uparrow}\rangle).
\end{equation}

ii)  Bell-operator measurement performed on the ``input'' particle and one
particle of the EPR pair. Bell operator has the following eigenstates:

\begin{eqnarray}
\label{Bell}
\nonumber |\Psi_-\rangle  = {1\over \sqrt2}(|{\uparrow}\rangle|{\downarrow}\rangle -
|{\downarrow}\rangle|{\uparrow}\rangle),
\\
\nonumber|\Psi_+\rangle = {1\over \sqrt2}(|{\uparrow}\rangle|{\downarrow}\rangle +
|{\downarrow}\rangle|{\uparrow}\rangle),
\\
|\Phi_-\rangle = {1\over \sqrt2}(|{\uparrow}\rangle|{\uparrow}\rangle -
|{\downarrow}\rangle|{\downarrow}\rangle),
\\
\nonumber |\Phi_+\rangle = {1\over \sqrt2}(|{\uparrow}\rangle|{\uparrow}\rangle +
|{\downarrow}\rangle|{\downarrow}\rangle).
\end{eqnarray}

iii) Transmission of the outcome of the Bell measurement and
appropriate unitary operation on the second particle of the EPR pair
(the ``output'' particle). The possible operations are: {\em nothing}
in the case of $|\Psi_-\rangle$ and $\pi$ rotations around $\hat x$,
$\hat y$, and $\hat z$ axes for the there other outcomes.

Completing (i)-(iii) ensures transportation of the pure state of the
input particle to the state of the output particle. It also ensures
transportation of correlations: if the input particle were correlated
to other systems, then the output particle ends up correlated to these
systems in the same way.

Unitary operations can be performed more or less effectively on all
systems. (The operations on  quantum states of a microwave cavity can be
performed indirectly through a manipulation
of  atomic  state and the interchange of the states of the atom and the
cavity.)  Preparation of the EPR pair is  also achievable (with various
levels of difficulty) for all systems. The main difficulty is
performing the Bell measurement.

\section{Bell-operator measurement without interaction between quantum
  systems}
\label{proof}

We shall prove here that it is impossible to perform complete
(nondegenerate) Bell-operator measurement without using interaction
between quantum systems.  We  allow  any unitary
transformation of single particle states and we are allowed to perform
any local single-particle measurement.

According to the standard (von Neumann) approach, the measurement
procedure can be divided into two stages: the unitary linear evolution
and local detection. There are four distinct (orthogonal)
single-particle states which are involved in the definition of the
Bell states: there are two channels, and a two-level system enters
into each channel.  We  name the channels left (L) and right (R)
corresponding to the way the Bell states (\ref{Bell}) are written,
i.e., in the explicit notation, $|\Psi_-\rangle =
(1/\sqrt2)(|{\uparrow}\rangle_L|{\downarrow}\rangle_R -
|{\downarrow}\rangle_L|{\uparrow}\rangle_R)$. The general form of the
unitary linear evolution of the teleportation procedure for the four
states can be written in the following form:
\begin{eqnarray}
\label{basic}
\nonumber |{\uparrow}\rangle_L \rightarrow \sum a_i |i\rangle,
\\
\nonumber |{\downarrow}\rangle_L \rightarrow \sum b_i |i\rangle,
\\
|{\uparrow}\rangle_R \rightarrow \sum c_i |i\rangle,
\\
\nonumber |{\downarrow}\rangle_R \rightarrow \sum d_i |i\rangle,
\end{eqnarray}
where $\{|i\rangle\}$ is a set of orthogonal single-particle local states.
The ``linearity'' means that the evolution of the particle in one
channel is independent on the state of the particle in the other
channel and therefore Eq. (\ref{basic}) is enough to define the evolution of the
Bell states:
\begin{eqnarray}
\label{uni}
\nonumber |\Psi_-\rangle    \rightarrow \sum_{i,j} \alpha_{ij} |i\rangle |j\rangle,
\\
\nonumber |\Psi_+\rangle \rightarrow \sum_{i,j} \beta_{ij}  |i\rangle |j\rangle,
\\
|\Phi_-\rangle \rightarrow \sum_{i,j} \gamma_{ij} |i\rangle |j\rangle,
\\
\nonumber |\Phi_+\rangle \rightarrow \sum_{i,j} \delta_{ij} |i\rangle |j\rangle.
\end{eqnarray}
In the right hand side the sum is on all different pairs $\{i,j\}$ and
the order is irrelevant. States of distinguishable particles
correspond to different $ |i\rangle$s. If the particles are identical,
$ |i\rangle |j\rangle$ signifies properly symmetrized states $
{(1/\sqrt 2 ) }(|i\rangle_1 |j\rangle_2 \pm |j\rangle_1 |i\rangle_2)$.

We assume that we have only local detectors, therefore, only the
product states $ |i\rangle |j\rangle$ (and not their superpositions)
can be detected.  Measurability of the nondegenerate Bell operator
means that there is at least one nonzero coefficient of every kind
$\alpha_{ij}, \beta_{ij}, \gamma_{ij}, \delta_{ij}$ and if, for a
certain $i,j$, it is not zero, then all others are zero.

If the particles entering the two channels are not identical, then
there are strong restrictions on the unitary evolution (\ref{basic})
which follows from the fact that the particles do not change their
identity:
\begin{eqnarray}
\label{nonid}
\nonumber a_i \neq 0 ~=>~c_i=d_i=0,
\\
\nonumber b_i \neq 0 ~=>~c_i=d_i=0,
\\
c_i \neq 0 ~=>~a_i=b_i=0,
\\
\nonumber d_i \neq 0 ~=>~a_i=b_i=0.
\end{eqnarray}
The equations for coefficients of the decomposition of the Bell states
after the unitary evolution are as follows.

 For $i=j$ we have
\begin{eqnarray}
\label{equaii}
\nonumber \alpha_{ii}  = a_i d_i - b_i c_i,
\\
\nonumber \beta_{ii}  = a_i  d_i + b_i c_i ,
\\
 \gamma_{ii}  = a_i  c_i - b_i d_i,
\\
\nonumber  \delta_{ii}    = a_i c_i +  b_i d_i,
\end{eqnarray}
but  from (\ref{nonid}) we immediately obtain  
\begin{equation}
\label{=0}
 \alpha_{ii}  = \beta_{ii}  =  \gamma_{ii}  = \delta_{ii} = 0 . 
\end{equation}

For  $i\neq j$ we have 
\begin{eqnarray}
\label{equa}
\nonumber \alpha_{ij}  &= a_i d_j + a_j d_i - (b_i c_j + b_j c_i),
\\
\nonumber \beta_{ij}  &= a_i d_j + a_j d_i + b_i c_j + b_j c_i,
\\
 \gamma_{ij}  &= a_i c_j + a_j c_i - (b_i d_j + b_j d_i),
\\
\nonumber  \delta_{ij}  &= a_i c_j + a_j c_i + b_i d_j + b_j d_i.
\end{eqnarray}
Let us assume that  $a_i\neq 0$ and $\alpha_{ij}\neq 0$. Then, taking
into account (\ref{nonid})  and the requirement of the measurability of the
Bell operator, we obtain
\begin{eqnarray}
\nonumber \alpha_{ij}  =  a_i d_j  - b_i c_j \neq 0,
\\
\nonumber \beta_{ij}  = a_i d_j  + b_i c_j = 0 ,
\\
 \gamma_{ij}  = a_i c_j - b_i d_j = 0,
\\
\nonumber \delta_{ij}  = a_i c_j  + b_i d_j = 0.
\end{eqnarray}
However, we can immediately see that this set of equations does not
have a solution. Since the equation are essentially symmetric relative
to  the various coefficients, we  get the same result while choosing
other nonzero coefficients. Therefore, we have proved the statement
for distinguishable particles.  

In fact, the proof yields more than the unmeasurability of the
nondegenerate Bell operator. We have proved that even a degenerate
Bell states operator measurement which separates just one Bell state
is impossible. Note, however, that degenerate Bell-state operator
measurement which distinguishes one pair of Bell states from the other
pair is possible.  For example, measurements of the $\sigma_{z1}$ and
$\sigma_{z2}$ distinguishes between the pair of states $|\Psi\rangle$
(for both of which the two outcomes are different) and the pair of
states $|\Phi\rangle$ (for which the two outcomes are identical).
Obviously, these are {\em demolition} measurements and we cannot
continue to measure the particles and specify all the Bell states.

Let us turn to the analysis of Bell measurement on identical particles
starting with bosons \cite{Norbert}. Taking into account
symmetrization, we obtain the equations (\ref{equaii}) and
(\ref{equa}) for the coefficients for different $i,j$ also for this
case, except for the
overall factor $1/\sqrt 2$ in equations (\ref{equa}) (we remind that
$|i\rangle|j\rangle$ means $ {(1/\sqrt 2 ) }(|i\rangle_1 |j\rangle_2 +
|j\rangle_1 |i\rangle_2)$).  For bosons, however, we have no
restrictions (\ref{nonid}) and therefore, we cannot claim immediately
that Eq. (\ref{=0}) holds.  The measurability of the nondegenerate
Bell operator requires that for any given $i$ at least three out of
the coefficients $ \alpha_{ii},~ \beta_{ii},~ \gamma_{ii}, ~
\delta_{ii} $ are zero. From equations (\ref{equaii}) it follows that
the fourth coefficient must be zero too and therefore we obtain Eq.
(\ref{=0}) also for the identical bosons. Thus, using (\ref{equaii})
again, we obtain
\begin{equation}
      a_i d_i =  b_i c_i = a_i  c_i =   b_i d_i =0.
\end{equation}
Therefore, at least two coefficients out of four, are zero: either  $
a_i = b_i=0$ or  $ c_i = d_i =0$.

Let us assume now $\alpha_{ij}\neq 0$ (and therefore $\beta_{ij} =
\gamma_{ij} = \delta_{ij} = 0$) and assume that $ a_i = b_i=0$. Then,
equations (\ref{equa}) become
\begin{eqnarray}
\nonumber  \alpha_{ij}  =  a_j d_i  -  b_j c_i \neq 0,
\\
\nonumber  \beta_{ij}  =  a_j d_i  +  b_j c_i = 0,
\\
 \gamma_{ij}  =  a_j c_i -  b_j d_i =0,
\\
\nonumber   \delta_{ij}  = a_j c_i + b_j d_i =0.
\end{eqnarray}
These equations, however, do not have a solution. It is easy to see
that also  for all other cases  there are no   solutions which proves the
statement for bosons. 

For bosons, in contrast with the case of non-identical particles, it is
possible to measure {\it degenerate} Bell operator which distinguishes
 two
Bell states. When we consider degenerate Bell operator, it is not true
anymore that at least three out of the coefficients $ \alpha_{ii},~
\beta_{ii},~ \gamma_{ii}, ~ \delta_{ii} $ are zero. A particular
solution \cite{Harold,BrMa,Mich} allows to distinguish two out
of four Bell states. The unitary linear transformation is
\begin{eqnarray}
\label{single}
\nonumber |{\uparrow}\rangle_L \rightarrow {1\over \sqrt2} (|1\rangle
+ |3\rangle),
\\
\nonumber |{\downarrow}\rangle_L \rightarrow  {1\over \sqrt2} (|2\rangle +|4\rangle), 
\\
|{\uparrow}\rangle_R \rightarrow  {1\over \sqrt2} (|1\rangle -|3\rangle),
\\
\nonumber |{\downarrow}\rangle_R \rightarrow  {1\over \sqrt2} (|2\rangle -|4\rangle),
\end{eqnarray}
and it leads to
\begin{eqnarray}
\label{degen}
\nonumber |\Psi_-\rangle    \rightarrow &{1\over \sqrt2} (|2\rangle |3\rangle
-|1\rangle |4\rangle),
\\
\nonumber |\Psi_+\rangle\rightarrow & {1\over \sqrt2} (|1\rangle |2\rangle
-|3\rangle |4\rangle),
\\
|\Phi_-\rangle\rightarrow  &{1\over 2} (|1\rangle |1\rangle -|3\rangle
|3\rangle -|2\rangle |2\rangle + |4\rangle |4\rangle),
\\
\nonumber |\Phi_+\rangle\rightarrow & {1\over 2} 
(|1\rangle |1\rangle -|3\rangle
|3\rangle + |2\rangle |2\rangle - |4\rangle |4\rangle).
\end{eqnarray}
(Note again that symmetrization is not written explicitly. For
example, $|3\rangle |4\rangle $ means $ {(1/\sqrt 2 ) }(|3\rangle_1
|4\rangle_2 + |4\rangle_1 |3\rangle_2)$.) This scheme can be realized
in a laboratory for photon polarization states using a beam-splitter
followed by polarizing beam-splitters and four detectors
\cite{Zei-trie}, see Fig.~1.

The most difficult case for the analysis is when the particles are
identical fermions. Due to   the (anti)symmetrization, Eq. (\ref{=0})
holds    for any choice of basic
unitary evolutions   (\ref{basic}) and therefore we do not
have equations analogous to (\ref{equaii}) but only  the analog of equations
(\ref{equa}):
\begin{eqnarray}
\label{equaf}
\nonumber \alpha_{ij}  &= a_i d_j - a_j d_i - (b_i c_j - b_j c_i),
\\
\nonumber \beta_{ij}  &= a_i d_j - a_j d_i + b_i c_j - b_j c_i,
\\
 \gamma_{ij}  &= a_i c_j - a_j c_i - (b_i d_j - b_j d_i),
\\
\nonumber  \delta_{ij}  &= a_i c_j - a_j c_i + b_i d_j - b_j d_i.
\end{eqnarray}
Again, measurability of the nondegenerate Bell operator means  that
there is at least one nonzero coefficient of every kind
$\alpha_{ij}, \beta_{ij},  \gamma_{ij},  \delta_{ij}$  
 and if, for a
certain $i,j$, it is not zero, then all others are zero.
Thus, we can obtain from (\ref{equaf}) the following equations:

If the detection of $|i\rangle|j\rangle$ signifies finding $|\Psi_-\rangle$ or
$|\Psi_+\rangle$, i.e., $\alpha_{ij}$ or  $\beta_{ij}$ are not zero, then
\begin{equation}
\label{psi}
{a_i\over a_j} ={c_i\over c_j} \neq {b_i\over b_j}= {d_i\over d_j} .
\end{equation}

\begin{center} \leavevmode \epsfbox{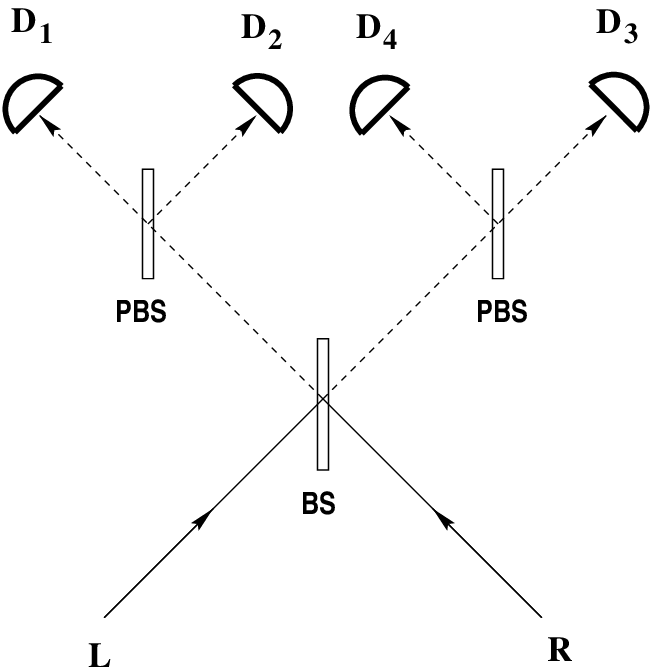} \end{center}

\noindent 
{\small {\bf Fig. 1.}  A scheme for the Bell operator measurement
  which distinguishes two out four Bell states.  The beam-splitter BS
  performs (up to irrelevant phases) interaction (\ref{single}), where
  for the input states: $|{\uparrow} \rangle \equiv |
  {\leftrightarrow} \rangle$, $|{\downarrow} \rangle \equiv
  |{\updownarrow} \rangle$, and for the output states: $|1 \rangle
  \equiv | {\leftrightarrow} \rangle _L$, $|2 \rangle \equiv
  |{\updownarrow} \rangle _L$, $|3 \rangle \equiv | {\leftrightarrow}
  \rangle _R$ $|4 \rangle \equiv |{\updownarrow} \rangle _R$. The
  polarization beam-splitters PBS transmit horizontal polarization and
  reflect vertical polarization. In this configuration each detector
  $D_i$ detects the state $|i\rangle$. The outcomes (2,3) and (1,4)
  signifies detecting of $|\Psi_-\rangle$, while (1,2) and (3,4)
  signifies detecting of $|\Psi_+\rangle$. The outcomes (1,1), (2,2),
  (3,3) and (4,4) correspond both to the detection of $|\Phi_-\rangle$
  and to the detection of $|\Phi_+\rangle$ which cannot be
  distinguished in this scheme.  }

\vskip .4cm

If the detection of $|i\rangle|j\rangle$ signifies finding $|\Phi_-\rangle$ or
$|\Phi_+\rangle$, i.e., $\gamma_{ij}$ or  $\delta_{ij}$ are not zero, then
\begin{equation}
\label{phi}
{a_i\over a_j} ={d_i\over d_j} \neq {b_i\over b_j}= {c_i\over c_j} .
\end{equation} 
The equations (\ref{psi}), (\ref{phi}) are valid  provided there are no
vanishing denominators.

Let us first prove that there cannot be a ``common'' state in the
product states corresponding to finding  $|\Psi\rangle$ ($+$ or $-$) and
$|\Phi\rangle$ ($+$ or $-$). Let us assume the opposite, that, say, 
 $|i\rangle|j\rangle$
corresponds to finding    $|\Psi\rangle$,  while  $|k\rangle|j\rangle$
corresponds to  finding $|\Phi\rangle$. 
 Then we have equations (\ref{psi}) as they are and
equation (\ref{phi}) with index $k$ instead of $i$:
\begin{equation}
\label{phi-kj}
{a_k\over a_j} ={d_k\over d_j} \neq {b_k\over b_j}= {c_k\over c_j} .
\end{equation} 
First, we will see that there cannot be zero in any of the
denominators in equations (\ref{psi}) and (\ref{phi-kj}), i.e., that none of the coefficients with index $j$
 vanish. From the fact that  for the indices $\{i, j\}$  exactly one equation out of
(\ref{equaf}) does not vanish follows that the two coefficients
of the same kind, e.g., $a_i,a_j$ cannot vanish
simultaneously. The same is true   for the indices $\{k, j\}$. 
  Since $|i\rangle|j\rangle$
corresponds to finding  $|\Psi\rangle$  we must have $\gamma_{ij}=
\delta_{ij}=0$ and $\alpha_{ij}\neq 0$ or   $\beta_{ij}\neq 0$.
Then, from (\ref{equaf}) follows that vanishing coefficients might
appear only in pairs:
$a_j=c_j=0$ or $b_j=d_j=0$. Similarly, since $|k\rangle|j\rangle$
corresponds to finding  $|\Phi\rangle$, we might have either 
$a_j=d_j=0$ or $b_j=c_j=0$. Therefore, if one coefficient vanishes then
all coefficients  vanish, which is impossible. Therefore, the
denominators in equations (\ref{psi}) and (\ref{phi-kj}) do not
vanish.
 We can always find an equality in 
 (\ref{psi}) which is not $0=0$ and we can divide  corresponding
 inequality in (\ref{phi-kj}) by the equality. We  obtain 
\begin{equation}
\label{=>phi}
{a_k\over a_i} \neq {c_k\over c_i}~~{\rm or}~~
{b_k\over b_i} \neq {d_k\over d_i}.
\end{equation}
Similarly, dividing the inequality from  (\ref{psi}) by the equality from
(\ref{phi})  leads to 
\begin{equation}
\label{=>psi}
{a_k\over a_i} \neq {d_k\over d_i}~~{\rm or}~~
{b_k\over b_i} \neq {c_k\over c_i} .
\end{equation}
From equation (\ref{=>phi}) follows that $|i\rangle|k\rangle$
corresponds to finding $|\Phi\rangle$ while equation (\ref{=>psi})
yields that $|i\rangle|k\rangle$ corresponds to finding
$|\Psi\rangle$. This contradiction ends the proof that it cannot be
that $|i\rangle|j\rangle$ corresponds to finding $|\Psi\rangle$, while
$|k\rangle|j\rangle$ corresponds to finding $|\Phi\rangle$.

 We have shown
that if detecting  $|i\rangle|j\rangle$ corresponds to finding
$|\Psi\rangle$  while detecting    $|k\rangle|m\rangle$  corresponds to finding
$|\Phi\rangle$ then all four states     $|i\rangle, |j\rangle,
|k\rangle, |m\rangle$  are different.
From  equation (\ref{psi}) and
equation (\ref{phi-kj}), with index $m$ instead of $j$, we always can
find an equality and the inequality for the same type of
coefficients. For example,
\begin{equation}
\label{ijmk}
{a_i\over a_j} ={c_i\over c_j}, ~~~~~ {a_k\over a_m} \neq {c_k\over
  c_m} .
\end{equation} 
Thus, there must be at least one out of the following
inequalities
\begin{equation}
\label{ijmk1}
{a_i\over a_m}  \neq{c_i\over c_m}, ~~{\rm or}~~~ {a_j\over a_k}
  \neq {c_j\over  c_k} .
\end{equation} 
Therefore, either $|i\rangle|m\rangle$ or $|k\rangle|j\rangle$
corresponds to detection of $|\Phi \rangle$. This  contradicts what we have proved above, since $|i\rangle|j\rangle$ corresponds to
detection of $|\Psi \rangle$. We have proved that also for fermions it
is impossible to measure nondegenerate Bell operator without
interaction between two quantum systems.

Note that nondegenerate Bell operator measurement which allows to
distinguish two out of four Bell states can be performed. In fact, it
can be done in the same way as with bosons. The single-particle
transformations (\ref{single})  leads (instead of (\ref{degen}) for
  bosons) to 
\begin{eqnarray}
\label{degenf}
\nonumber |\Psi_+\rangle &\rightarrow {1\over \sqrt2} (|1\rangle |2\rangle
-|3\rangle |4\rangle) ,
\\
\nonumber |\Psi_-\rangle    &\rightarrow {1\over \sqrt2} (|3\rangle |2\rangle
-|1\rangle |4\rangle) ,
\\
 |\Phi_-\rangle &\rightarrow {1\over \sqrt2} ( -|1\rangle |3\rangle -
|2\rangle |4\rangle) ,
\\
\nonumber |\Phi_+\rangle &\rightarrow {1\over \sqrt 2} (- |1\rangle
|3\rangle + |2\rangle |4\rangle) .
\end{eqnarray}
(Again, the symmetrization is not written explicitly, e.g., $|3\rangle
|4\rangle $ means $ {(1/\sqrt 2 ) }(|3\rangle_1 |4\rangle_2 -
|4\rangle_1 |3\rangle_2)$. We see that finding in the final
measurement $|1\rangle |2\rangle$ or $|3\rangle |4\rangle$ signifies
detecting of $|\Psi_-\rangle$ while finding in $|3\rangle |2\rangle$
or $|1\rangle |4\rangle$ signifies detecting of $|\Psi_+\rangle$.
States $|1\rangle |3\rangle$ and $|2\rangle |4\rangle$ correspond to both
$|\Phi_-\rangle$ and $|\Phi_+\rangle$ which cannot be distinguished in
this scheme. (Slight modification of this method is required for
distinguishing any other two out of four Bell states.)

\section{Teleportation using  interaction between quantum
  systems}
\label{tele-int}

If we allow interaction between quantum particles we can achieve
reliable (100\% efficient teleportation). In this case we can perform
measurement of the nondegenerate Bell operator, or we can use the method
of ``crossed'' nonlocal measurements \cite{tele-V}.

For measurement of the Bell operator consider the 
interaction between the particles, say, according to the following
rule:
\begin{eqnarray}
\label{intera}
\nonumber
|{\uparrow}\rangle|{\uparrow}\rangle \rightarrow 
|{\uparrow}\rangle|{\downarrow}\rangle ,
\\
\nonumber
|{\uparrow}\rangle|{\downarrow}\rangle \rightarrow 
|{\uparrow}\rangle|{\uparrow}\rangle,
\\
|{\downarrow}\rangle|{\uparrow}\rangle \rightarrow 
|{\downarrow}\rangle|{\uparrow}\rangle, 
\\
\nonumber
|{\downarrow}\rangle|{\downarrow}\rangle \rightarrow 
|{\downarrow}\rangle|{\downarrow}\rangle.
\end{eqnarray}
This interaction is a ``conditional spin flip''. It transforms the Bell states (\ref{Bell}) to product
states
\begin{eqnarray}
\label{prod}
\nonumber |\Psi_-\rangle    \rightarrow 
 {1\over \sqrt2}(|{\uparrow}\rangle - |{\downarrow}\rangle )|{\uparrow}\rangle,
\\
\nonumber |\Psi_+\rangle \rightarrow 
{1\over \sqrt2}(|{\uparrow}\rangle + |{\downarrow}\rangle )|{\uparrow}\rangle,
\\
|\Phi_-\rangle \rightarrow {1\over \sqrt2}(|{\uparrow}\rangle -
|{\downarrow}\rangle )|{\downarrow}\rangle,
\\
\nonumber |\Phi_+\rangle \rightarrow {1\over \sqrt2}(|{\uparrow}\rangle +
|{\downarrow}\rangle )|{\downarrow}\rangle,
\end{eqnarray}
which can be measured by local detectors. Note that the conditional
spin flip 
(\ref{intera}) is equivalent (in another bases) to the conditional
phase flip:
\begin{eqnarray}
\label{pha-fli}
\nonumber
|{\uparrow}\rangle|{\uparrow}\rangle \rightarrow 
|{\uparrow}\rangle|{\uparrow}\rangle,
\\
\nonumber
|{\uparrow}\rangle|{\downarrow}\rangle \rightarrow -
|{\uparrow}\rangle|{\downarrow}\rangle,
\\
|{\downarrow}\rangle|{\uparrow}\rangle \rightarrow 
|{\downarrow}\rangle|{\uparrow}\rangle,  
\\
\nonumber
|{\downarrow}\rangle|{\downarrow}\rangle \rightarrow 
|{\downarrow}\rangle|{\downarrow}\rangle.
\end{eqnarray}
This is so when we use the $\hat x$ direction (instead of $\hat z$)
for the spin of the second particle. This interaction is nonlinear in
a sense that one quantum system changes its state depending on the
state of  another quantum system.

In a slightly different method for Bell measurement, the two particles
do not interact one with the other but both interact with an auxiliary
quantum particle. Consider consecutive interactions of the two
particles with a spin$-{1\over2}$ particle prepared initially in the
state $|{\uparrow}\rangle$.  Each interaction is described by
(\ref{intera}) when the second terms in the products signify the spin
states of the auxiliary particle.  This operation allows to distinguish
between $|\Psi \rangle$ and $|\Phi \rangle$ states. Indeed, each of  the $|\Psi
\rangle$ states leads to the flip of the spin of the auxiliary
particle, while each of the $|\Phi\rangle$ states does not. In order to distinguish
between $|\Psi_-\rangle$ and $|\Psi_+\rangle$ (or between
$|\Phi_-\rangle$ and $|\Phi_+\rangle$) we have to repeat the
procedure, i.e. to perform the two consecutive measurements
(\ref{intera}) but now in the spin-$x$ component bases.

Another way to view the measurement procedure for Bell measurement
described above is to consider it as consecutive measurements of two
two-particle variables:
\begin{eqnarray}
\label{nonl}
\nonumber
({\sigma_1}_z  +{\sigma_2}_z){\rm mod4},\\
({\sigma_1}_x +
{\sigma_2}_x){\rm mod4} .
\end{eqnarray}
(The spin components are measured in the units of $\hbar \over 2$.)
These two-particle variables are ``local'' since both particle 1 and
particle 2 are located at the same site, but these variables can be
measured even if the particles located in spatially separated sites, in which case they called {\em nonlocal} measurements \cite{AAV86}.  A
modification of nonlocal measurements when we ``cross'' the
interactions with the two particles in time,  leads to an
alternative method of teleportation \cite{tele-V}.  In order to
teleport a quantum state from particle 1 to particle 2 and, at the
same time, the quantum state of particle 2 to particle 1, the following
(nonlocal in space-time) variables  should be measured (see Fig. 2):
\begin{eqnarray}
\label{swap}
\nonumber
{\cal Z} \equiv  \Bigl({\sigma_1}_z(t_1) + {\sigma_2}_z(t_2)\Bigr){\rm mod4}, \\
{\cal X} \equiv \Bigl({\sigma_1}_x(t_2) + {\sigma_2}_x(t_1)\Bigr){\rm
mod4}.
\end{eqnarray}
For any set of outcomes of the nonlocal
measurements (\ref{swap}) the spin state is teleported; in some cases
the state is rotated by $\pi$ around one of the axes, but the resulting rotation can be inferred from
the nonlocal measurements. We can complete, then, the teleportation by
the following transformations
\begin{eqnarray}
\label{table}
\nonumber
(\cal Z, \cal X )~~~~~~~~~ \pi~ {\rm rotation~}\\
\nonumber (0,0) ~~~~~~~~~~~ y ~{\rm axis}~~~~~ \\
 (2,0) ~~~~~~~~~~~ x ~{\rm axis}~~~~~ \\
\nonumber (0,2) ~~~~~~~~~~~ z ~{\rm axis}~~~~~ \\
\nonumber (2,2) ~~~~~~~~~ {\rm no~ rotation}
\end{eqnarray}

In order to perform nonlocal measurements (\ref{nonl}) or
(\ref{swap}), correlated pairs of auxiliary particles located in the
sites of particle 1 and 2 are required. For example, for measuring
$({\sigma_1}_z +{\sigma_2}_z){\rm mod4}$ a pair of spin-$1\over2$
particles in a singlet state is used. The local interaction
(\ref{intera}) in each site between the particle in and the auxiliary
particle from the singlet is performed (the second ``ket''s in the 
product signifies the state of the auxiliary particle). Then

\begin{center} \leavevmode \epsfbox{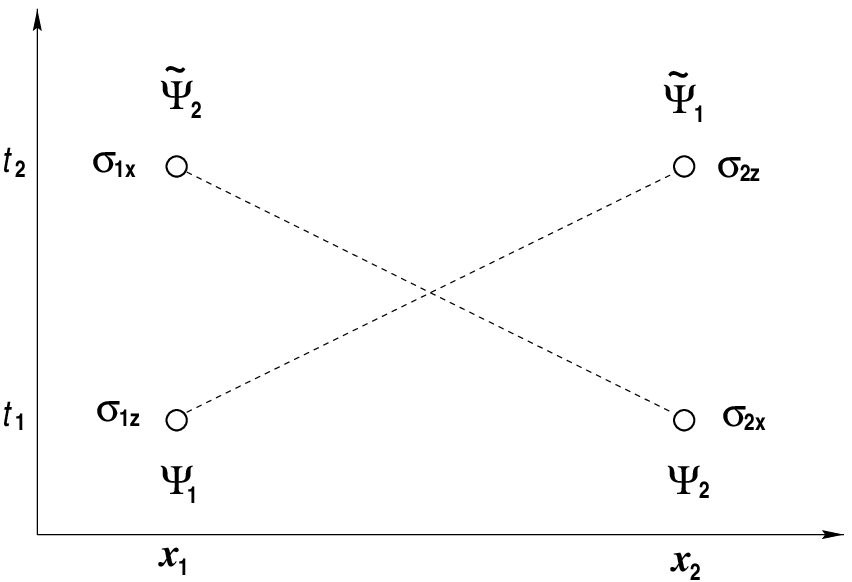} \end{center}

\noindent 
{\small {\bf Fig. 2.} Space-time diagram of ``crossed'' nonlocal
  measurements which result in two-way teleportation, i.e., swapping
  of the spin states. Space-time locations of local couplings are
  shown.  When the nonlocal measurements (\ref{swap}) are completed,
  the states of the two particles are interchanged up to local $\pi$
  rotations, $\tilde \Psi_i$ signifies $\Psi_i$ rotated according to
  table  (\ref{table}).  }

\vskip .4cm

\noindent
${\sigma}_z$ measurement is performed on the auxiliary particles in
both sites. If the outcomes are the same it means that just one 
auxiliary particle has been flipped, i.e., ${\sigma_1}_z$ and
${\sigma_2}_z$ are different and therefore $({\sigma_1}_z
+{\sigma_2}_z){\rm mod4} = 0$.  If the outcomes are different, then
either both or none of the auxiliary particles have been flipped and
therefore $({\sigma_1}_z +{\sigma_2}_z){\rm mod4} = 2$.

\section{Towards experimental realization of reliable  teleportation }
\label{tele-exp}

In recent experiments, which announced first implementation of
teleportation, the polarization state has been transported from one
photon to another. The experiments performed in Innsbruck, pure state
teleportation \cite{Inn} and correlated state teleportation
\cite{Inn-mix}, have (theoretical) success rate of 25\%. A feasible
modification of the experimental setup for the Bell-state analyzer
(bringing it to the one described in Fig. 1) can increase the success
rate of these experiments to 50\%. However, due to lack of an effective
photon-photon interaction, these experiments cannot be modified to
reach 100\% success rate, as it was proved in Section \ref{proof}.
Popescu \cite{Pop} has found an ingenious way to overcome this
difficulty  by using polarization and location degrees of
freedom of the {\it same} photon. Thus, the Rome experiment
\cite{Rome} which implemented his idea has (theoretically) 100\%
success rate. Unfortunately, this method works only for transportation
of the polarization state prepared on a particle which is the local
member of the EPR pair constituting the quantum channel between the
two sites.  It is not applicable for teleportation of an unknown state
of an external particle.
 
Thus, it seems that the most promising candidates for teleportation
experiment which might have 100\% success rate are proposals which
involve atoms and electro-magnetic cavities. First suggestions for
such experiments were made shortly after publication of the original
teleportation paper \cite{Slea,David,Cirac} and numerous modifications
appeared since.  The reason why implementation of these proposals
seems to be feasible is that in this case ``quantum-quantum''
interaction between the system carrying the quantum state and a system
from the EPR pair exists.  A dispersive interaction (DI) of a Rydberg
atom passing through a properly tuned micro-wave cavity leads to a
conditional phase flip as in Eq. (\ref{pha-fli}) depending on the
presence of a photon in the cavity. A resonant interaction (RI-$\pi$)
between the Rydberg atom and the cavity allows swapping of quantum
states of the atom and the cavity. Thus, manipulation of the quantum
state of the cavity can be achieved via manipulation of the state of
the Rydberg atom. The atom state is transformed by sending it through
appropriately tuned microwave zone. Moreover, the direct analog of
conditional spin flip interaction (\ref{intera}) can be achieved
through the Raman atom-cavity-field interaction \cite{ZG97}. No
teleportation experiment has been performed yet using these methods,
but it seems that the technology is not too far from this goal. About
the progress in this direction one can learn from recent experiments
on atom-cavity interactions \cite{Har}.

  Until further progress in technology it is
 not easy to predict which proposal will be implemented first.
 Assuming that resonant interactions between  atoms and the cavity can be
 performed  with a very good precision and that dispersive
 interaction is available with reasonable precision, the simplest way
 is to use the quantum channel consisting of a cavity and a
 Rydberg atom in a correlated state.  
 A particular resonant  interaction,   RI-$\pi$,   of 
 an excited atom passing through  an empty cavity,
 \begin{equation}
   \label{resona}
\nonumber
|e\rangle |0\rangle \rightarrow  {1\over \sqrt2}(|g\rangle |1\rangle +
|e\rangle |0\rangle)  ,
 \end{equation}
  prepares this
 quantum channel.

 The quantum
 state to be teleported is the state of another Rydberg atom. The Bell
 measurement is then performed on this atom and the cavity. An
 interaction between the atom and the cavity, the conditional phase
 flip via dispersive interaction, DI,
\begin{eqnarray}
\label{pha-fli-ca}
\nonumber
|e\rangle|0\rangle \rightarrow 
|e\rangle|0\rangle,
\\
\nonumber
|e\rangle|1\rangle \rightarrow -
|e\rangle|1\rangle,
\\
|g\rangle|0\rangle \rightarrow 
|g\rangle|0\rangle,
\\
\nonumber
|g\rangle|1\rangle \rightarrow 
|g\rangle|1\rangle,
\end{eqnarray}
 disentangles the the following Bell states:
\begin{eqnarray}
\label{Bell-ca}
\nonumber |\Psi_{\pm}\rangle  = {1\over 2}\Bigl(|e\rangle (|0\rangle
-|1\rangle)  \pm |g\rangle (| 1\rangle +|0\rangle\Bigr),
\\
 |\Phi_{\pm}\rangle  = {1\over 2}\Bigl(|e\rangle (|0\rangle
+|1\rangle)  \pm |g\rangle (| 1\rangle -|0\rangle\Bigr).
\end{eqnarray}
The Bell states (\ref{Bell-ca})  have the form of Eq. \ref{Bell} when the first
$|{\uparrow}\rangle$ in the product  is
identified with $|e\rangle$, the second $|{\uparrow}\rangle$,  with
 $(1/\sqrt2)(|0\rangle +
|1\rangle)$, etc.  Measurement of the atom state and the cavity state
 completes the Bell measurement procedure.

In order to make the measurement of the cavity state we perform another
  resonant interaction, RI-$\pi$/2,  between the cavity an auxiliary
  atom prepared initially in the ground
state,   
 \begin{eqnarray}
   \label{resona1}
\nonumber
|g\rangle |1\rangle \rightarrow  |e\rangle |0\rangle, \\
 |g\rangle |0\rangle \rightarrow  |g\rangle |0\rangle. 
 \end{eqnarray}
This interaction transfers the quantum
state of the cavity to this atom. The final measurements on the atoms
 distinguish between the states
\begin{equation}
  \label{atom-sta}
  {1\over \sqrt2}(|g\rangle +
|e\rangle), ~~~ {1\over \sqrt2}(|g\rangle - 
|e\rangle).
\end{equation}
Since detectors can distinguish between $|g\rangle$ and $|e\rangle$,
the atoms should rotate their states passing through the appropriate
microwave zones before the detection. When the Bell measurement is
completed, the quantum state is teleported up to the known local
transformation determined by the results of the Bell measurement. The
scheme for this teleportation procedure, apart from the final local
transformation, is presented in Fig. 3.

Another proposal for teleportation involving single cavity \cite{ZG97}
is based on Raman interaction. This scheme does not require microwave
zone interactions, but it requires additional atom for preparing the
initial state of the cavity, $(1/\sqrt2)(|0\rangle + |1\rangle)$.  A
very recent single-cavity teleportation proposal \cite{Almei}, which
employs three-particle (cavity and two atoms) correlated state, seems
to be more complicated.  The main advantage of this scheme, that it
does not require measurement of the nondegenerate Bell operator is not
very important since quantum-quantum interactions are available and,
consequently, the Bell measurement is not too problematic. On the
other hand, the complications related to the preparation of the three-particle
entanglement are significant.

One of the relatively simple methods for ``two-way'' teleportation,
i.e., swapping of quantum states of two separated atoms, is a direct
implementation of crossed nonlocal-measurement procedure presented in
the previous section.  Two pairs of correlated micro-wave cavities are
prepared as the two quantum channels. Preparation of such a channel
requires consecutive resonant interaction of an auxiliary atom,
prepared in the excited state: first, RI-$\pi$/2 with the cavity in
one site and second, RI-$\pi$ with the cavity in the other site, see
Fig. 4a.
 
 In the second step, the atom in each site passes, interacting
 dispersively, through the cavities in the ``crossed'' order, see Fig.
 4b. The dispersive interaction is given by Eq. (\ref{pha-fli-ca}) and
 it corresponds to measurement of $\sigma_z$, because $|{\uparrow}_z
 \rangle$ flips the states of the cavity defined as
\begin{equation}
  \label{basis}
|{\uparrow} \rangle \equiv {1\over
  \sqrt2}(|0\rangle + |1\rangle), ~~|{\downarrow} \rangle \equiv
{1\over \sqrt2}(|0\rangle - |1\rangle).
\end{equation}
This is the basis of the Bell states (\ref{Bell-ca}).  Since we also
have to make coupling to $\sigma_x$ we have to rotate the state of the
atom such that ``$x$'' rotates to ``$z$'' before coupling with the
cavity and rotate it back after the interaction.

\begin{center} \leavevmode \epsfbox{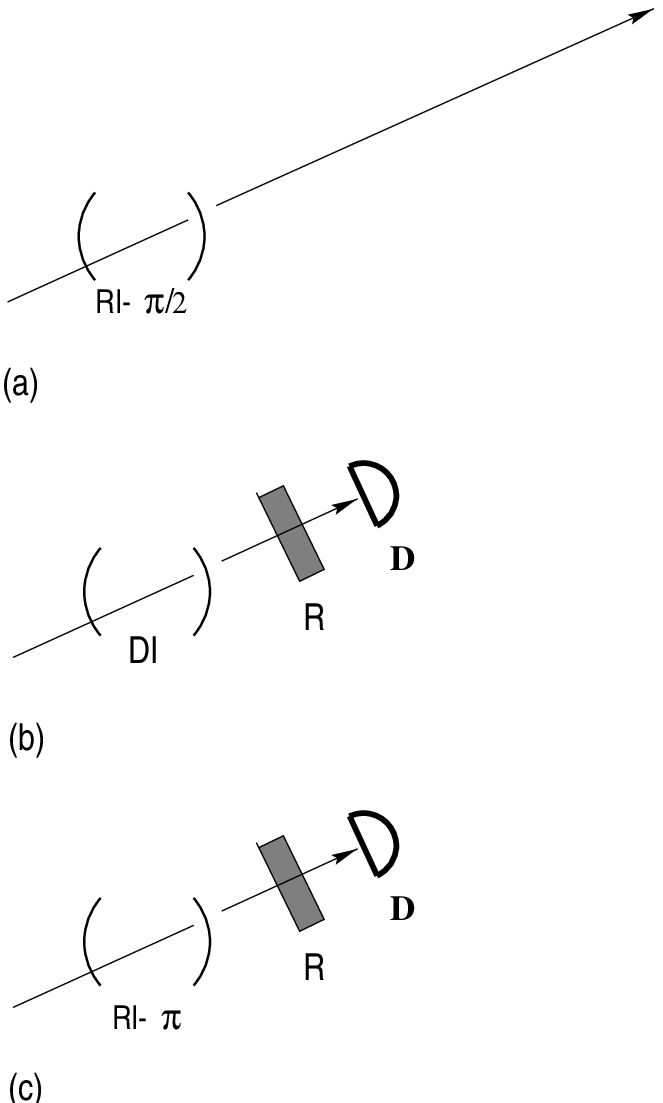} \end{center}

\noindent 
{\small {\bf Fig. 3.}  Single-cavity teleportation of a quantum state of an
  atom. The final stage, the rotation of the state according to
  the  results of the Bell measurement is not shown.
\\
  (a) Preparation of the quantum channel. An atom passes through a
  cavity and sent to a remote site.  Resonant interaction
  RI-$\pi$/2 given by (\ref{resona}) of  the atom
  in the cavity  creates the correlation.
  \\
  (b) The atom, carrying the quantum state to be teleported, passes through
  the cavity, microwave field zone R and is detected by the detector D. 
    Dispersive atom-cavity interaction DI given by
  (\ref{pha-fli-ca}) disentangles
  the Bell states (\ref{Bell-ca}). Microwave field zone R rotates the
  atom states (\ref{atom-sta}) to $|g\rangle$ and $ |e\rangle$
  accordingly, which are then distinguished by the detector D.
  \\
  (c) An auxiliary atom passes through the cavity in order to measure
  its state.  Resonant interaction  RI-$\pi$ transfers the
  states of the cavity to the atom. Then, the atom states are rotated
  in the microwave zone R and distinguished by the detector D.
}

\break

\begin{center} \leavevmode \epsfbox{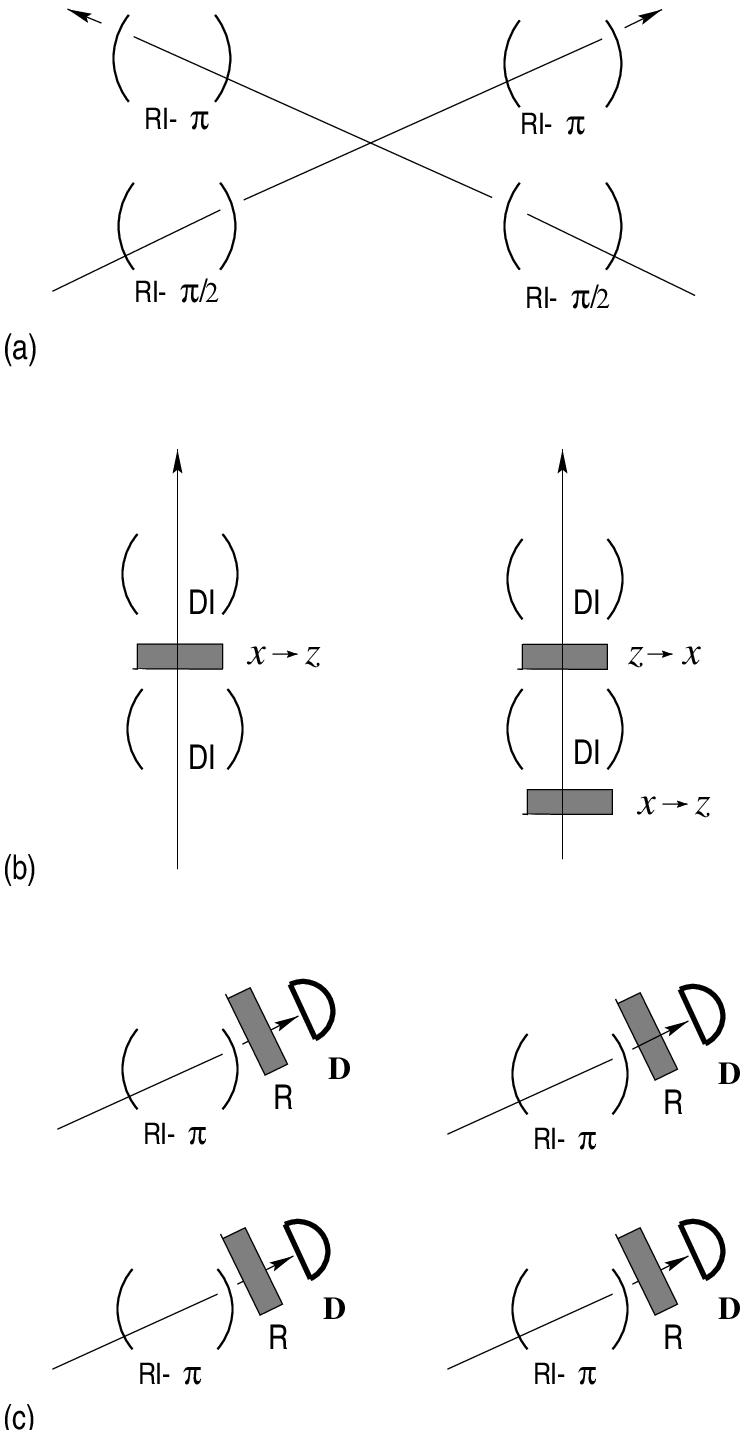} \end{center}

\noindent 
{\small{\bf Fig. 4.}  Two-way teleportation of quantum states of
  atoms.  The final rotations of the states of the atoms according to
  the results of nonlocal measurements are not shown.
  \\
  (a) Preparation of the quantum channels. Atom in the excited state
  makes resonant interaction RI-$\pi$/2 with a cavity in one site and
  resonant interaction RI-$\pi$ with corresponding cavity in another
  site.  Another atom makes the same interactions with the second pair
  of cavities. Each pair ends up in a correlated state $ (1/
  \sqrt2)(|0\rangle |1\rangle + |1\rangle |0\rangle)$.
  \\
  (b) Coupling of the atoms with the quantum channels.  An atom on
  each site passes through two cavities making dispersive interactions
  DI. In between, the atoms pass through microwave zones which make
  the appropriate rotations of their quantum states.
  \\
  (c) Local measurements of the quantum states of the cavities which
  yield the results of the nonlocal measurements.  Auxiliary atoms
  pass through cavities in order to measure their states.  Resonant
  interactions RI-$\pi$ transfer the states of the cavities to the
  atoms. Then, the atom states are rotated in the microwave zones R
  and distinguished by detectors D.  } \vskip .4cm

The third step, described in Fig. 4c, is the measurement of the states
of the cavities in the basis (\ref{basis}).  This stage, requires an
auxiliary atom for each cavity. The resonant interaction transfers the
correlated state of the pairs of cavities to the correlated states of
the atoms. Then the atoms pass through microwave zones  rotating
their states before detection by the  detectors which distinguish between
 ground and excited states. These outcomes of these local measurements
 yield the result of the
 nonlocal measurements (\ref{swap}) which determine the final
 transformation to be performed on the atoms (to be combined with $z
 \rightarrow x$ rotation for the first atom which was left out, see
 Fig. 4b.). 
Completing all stages of the procedure described above results in
two-way teleportation.

Note another (seemingly less economical) proposal for a two-way
teleportation of atom states using interaction with cavities
\cite{Mous-swa}. It essentially doubles one of the original atom
-cavity teleportation schemes \cite{David} reusing atoms which bring
the quantum states 
for receiving the quantum states from the other site after ``stripping'' from them  quantum information. Anyway, today
  the two-way  teleportation is still a gedanken experiment, at
least until one-way teleportation will be implemented.

One difficulty with teleportation of atomic states which should be
mentioned is that  usual experiments are performed with atomic {\em
  beams} and not with individual atoms. Such experiments might be good
for demonstration and studying experimental difficulties for
teleportation, but they cannot be considered as implementation of the
original wisdom of teleportation or used for cryptographic
purposes. In fact, optical experiments have this difficulty too, unless
  ``single-photon'' guns will be used. Both for atoms and in the optical
 regime this problem does not seem to be unsolvable, but it certainly
 brings attention to experiments with trapped ions, the experiments
 which individual quantum systems. There are many
 similarities between available manipulations with atoms and with ions
 so the methods discussed above  might be implemented for ion systems too.

\section{Teleportation of continuous variables}
\label{tele-cont}

In the framework of nonlocal measurements there is a natural way of
extending the teleportation scheme to the systems with continuous
variables \cite{tele-V}.  Consider two similar systems located far
away from each other and described by continuous variables $q_1$ and 
$q_2$ with corresponding conjugate momenta $p_1$ and $p_2$. In order
to teleport the quantum state of the first particle    $\Psi_1(q_1)$ to
the second particle (and the state of the second particle $\Psi_2(q_2)$ to the first) we perform the following
``crossed'' nonlocal measurements (see Fig. 5), obtaining the
outcomes $a$ and $b$:

\begin{center} \leavevmode \epsfbox{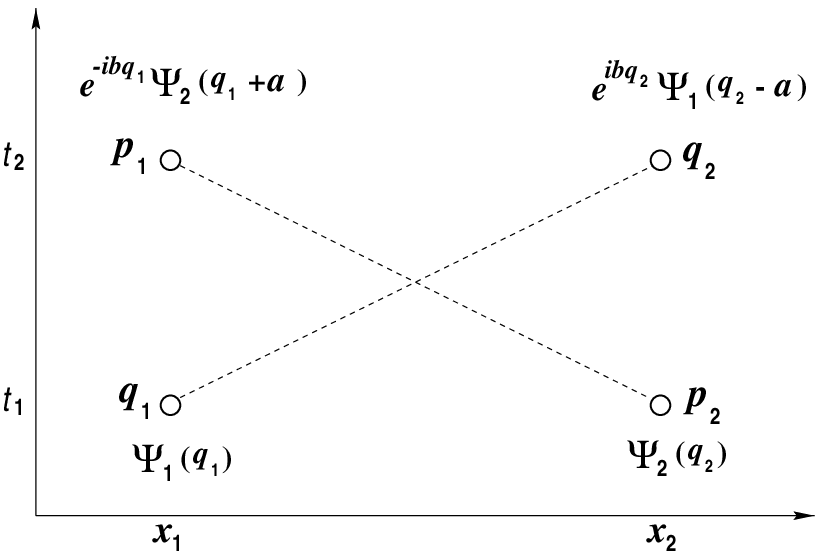} \end{center}

\noindent 
{\small {\bf \bf Fig. 5.} Space-time diagram of ``crossed'' nonlocal
  measurements which result in two-way teleportation of quantum states
  of quantum systems with continuous variables.  Space-time locations of
  local couplings are shown.  When the nonlocal measurements
  (\ref{cross-conti}) are completed, the states of the two particles
  are interchanged up to the known shifts in $q$ and $p$.  } \vskip .4cm

 \begin{eqnarray}
\label{cross-conti}
\nonumber
q_1(t_1) - q_2(t_2) = a,\\
 p_1(t_2) - p_2(t_1) = b. 
\end{eqnarray}

In Ref. \cite{tele-V} it is shown that these nonlocal ``crossed''
measurements ``swap'' the quantum states of the two particles up to
the known shifts in $q$ and $p$. Indeed, the  states of the particles
after completion of 
the measurements (\ref{cross-conti}) are
\begin{eqnarray}
\label{psi-shif}
\nonumber
\Psi_{f}(q_1)= e^{ibq_1} \Psi_2(q_1+a),\\
 \Psi_{f}(q_2)= e^{-ibq_2} \Psi_1(q_1-a) . 
\end{eqnarray}
The state of particle 2 after $t_2$, is the initial state of the particle
1 shifted by $-a$ in $q$ and  by $-b$ in $p$.  Similarly, the state of
particle 1 is the initial state of particle 2 shifted by $a$ in $q$
and by $b$ in $p$.  After transmitting the results of the local
measurements, $a$ and $b$, the shifts can be corrected by appropriate
kicks and back shifts (even if the quantum state is unknown), thus
completing a reliable teleportation of the state $\Psi_1(q_1)$ to
$\Psi_1(q_2)$ and of the state $\Psi_2(q_2)$ to $\Psi_2(q_1)$.

In order to perform nonlocal measurements of continuous variables
(\ref{cross-conti}) a quantum channel is required. While for the case
of spin-$1\over2$ particles the quantum channel was an EPR(-Bohm)
pair, for continuous variables the entangled state of the original EPR paper is
required:
\begin{eqnarray}
\label{EPR-Q}
\nonumber
Q_1 + Q_2 = 0,\\
 P_1 - P_2  = 0,
 \end{eqnarray}
 where $Q_1$ and $Q_2$ are continuous variables of the pair of
 auxiliary particles with corresponding conjugate momenta $P_1$ and $
 P_2$.
Two local von Neumann type interactions which can be written in the
interaction  Hamiltonian
\begin{equation}
\label{int-Q} 
H = g(t-t_1) P_1 q_1  - g(t-t_2) P_2 q_2,
\end{equation}
with normalized and localized around zero $g(t)$ lead to final state
of auxiliary particles such that 
\begin{equation}
\label{Q-final}
Q_{1f} + Q_{2f} = q_1(t_1) - q_2(t_2). 
 \end{equation}
Thus, local measurements of $Q_{1f}$ and $ Q_{2f}$ yield $a$ (of Eq. (\ref{cross-conti})). Another
EPR pair is needed for measuring $b$.

A generalization of the BBCJPW scheme to the case of continuous
variables is also possible. The quantum channel is again the EPR state
(\ref{EPR-Q}). Now, an analog of Bell measurement for the systems with
variables $q$ which carries the quantum state and the auxiliary
system with variable $Q_1$ is needed.  While it is not  easy to make
generalization of the Bell operator through definition of its
eigenstates, it is straightforward to generalize Bell operator
measurement from the two consecutive nonlocal measurements
(\ref{nonl}) applied in the spin-$1\over2$ particles case, to the following two
consecutive nonlocal measurements
 \begin{eqnarray}
\label{Bell-cont}
\nonumber
q + Q_1,\\ p - P_1.
\end{eqnarray}
These measurements ends up in finding one of the  ``shifted'' EPR states,
where the shifts are the outcomes of the measurements:
 \begin{eqnarray}
\label{EPR-shi}
\nonumber 
q + Q_1 = a,\\
  p - P_1 = b.
\end{eqnarray}  
At the end of this measurement the quantum state $\Psi(q)$ is
teleported, again up to the known shifts, to the remote particle of the EPR
pair, the state of which becomes 
\begin{equation}
\label{Q-final1}
\Psi_{f}(Q_2)= e^{ibQ_2} \Psi(Q_2+a) .
 \end{equation}
 The final stage of this teleportation procedure consist of the
 appropriate back shifts of the state in $Q_2$ and $P_2$ which result
 in transporting the quantum state $\Psi(q)$ in site 1 to the same
 quantum state $\Psi(Q_2)$ of a system in site 2. Of course, this
 method, unlike the crossed measurements scheme, yields only one-way
 teleportation. Both methods transport not just pure states but also
 correlations, when the input particle is not in a pure state before
 teleportation.

 Surprisingly, reliable teleportation of continuous variables seems to
 be possible to implement in a real laboratory. Braunstein and Kimble
 made a realistic proposal for teleporting quantum state of a single
 mode of the electro-magnetic field \cite{BK}. This
 remarkable result is an implementation of the scheme described in the
 previous paragraph.  In their methods $q$ is ``$x$''defined for a
 single mode of an elector-magnetic field, and correspondingly $p$ is
 the conjugate momentum of $x$. The analog of the EPR state
 (\ref{EPR-Q}) is obtained by shining squeezed light with certain
 $x$ from one side and squeezed light with certain $p$ from the other
 side of a simple beam splitter. The analog of measurements
 (\ref{Bell-cont}) is achieved using another simple beam splitter and  homodyne detectors. The shifts
 in $x$ and $p$ which complete the teleportation procedure can be done
  by combining the output field with coherent state of an
 appropriate amplitudes fixed by the results of the homodyne
 measurements. Preliminary experimental results for such 
 teleportation procedure has been obtained \cite{Br-priv}.
 Note also a very recent proposal \cite{Mol} for teleporting of a
 single-photon wave packet.

\section{Conclusions}
\label{conc}

We have shown  that without ``quantum-quantum''
interaction one cannot perform complete measurement of Bell
nondegenerate operator and therefore, one cannot perform reliable
teleportation of photon polarization states  using
 method employed in recent experiments, i.e., experiments without
 quantum-quantum interactions. Therefore, another
methods for teleportation have to be developed. Today's technology
allows quantum-quantum interaction between atoms and 
electro-magnetic  field in cavities which makes teleportation schemes using
such elements good candidates for implementation reliable
teleportation. Various schemes are briefly analyzed, in particular,
using the language of nonlocal measurement which proved to be helpful
for such problems. Schemes for one- and two-way teleportation which
seems to be easiest for implementation are proposed.

We have discussed methods for teleportation of system described by
continuous variables. One may see an apparent contradiction between
the proof of Section \ref{proof} that 100\% efficient teleportation
cannot be achieved using linear elements and single-particle state
detectors and the result of Braunstein and Kimble presented at the end
of the last section according to which one can perform a {\em
  reliable} teleportation of quantum state of a system with continuous
variables using beam-splitters and local measuring devices.  Indeed,
it is natural to assume that if reliable teleportation of a quantum
state of a two-level system is impossible at certain circumstances, it
is certainly impossible for quantum states of systems with continuous
variables.  However, although it is not immediately obvious, the
circumstances are very different.  There are several differences.  The
analog of Bell operator for continuous variable does not have among
its eigenvalues four states of the general form (\ref{Bell}) where
$|{\uparrow}\rangle$ and $|{\downarrow}\rangle$ signify some
orthogonal states.  Even more importantly, the beam-splitter, a simple
half-silvered mirror is {\em not} a linear element in the sense of
Section \ref{proof}. If we do not send any ``light'' from one side of
the beam-splitter, we still get the vacuum field from this port.  The
beam-splitter in Braunstein-Kimble experiment leads to
``quantum-quantum'' interaction: the variable $x$ of one of the output
ports of the beam splitter becomes equal to ${1\over \sqrt 2}(x_1 +
x_2)$, essentially, the sum of the quantum variables of the input
port. The absence of such ``quantum-quantum'' interactions is an
essential ingredient in the proof of Section \ref{proof}. The
beam-splitter, however, is linear for photons, but the homodyne
detectors which measure $x$ are not single-particle detectors for
photons -- using single-particle measuring devices is another
constraint used in the proof. The Braunstein-Kimble method is not
applicable directly for teleporting $\Psi (x)$ where $x$ is a spatial
position of a quantum system. Additional quantum-quantum interaction
which converts continuous variable of a real particle to the abstract
(although measurable) variable $x$ of their method is required.

The fact that Braunstein-Kimble method does not contradict the proof
does not make the method less interesting. The task is to find feasible
proposals for teleportation. The reason why linear devices were
considered in the proof is because  usually it is more difficult to
implement nonlinear interactions in a laboratory. Half-silvered
mirror, even if it can be considered as a nonlinear device in some
sense, can be easily used in a laboratory.
 (The main experimental difficulty of the
Braunstein-Kimble proposal is preparation of highly squeezed light,
which is necessary for high-efficiency teleportation.) On the other
hand, the fact that there are realistic proposals for teleportation
which do not fulfill the assumption of the proof of Section
\ref{proof} does not make the proof irrelevant. It still limits
wide class of teleportation proposals, in
particular those  in  which  quantum states are encoded in
photon polarization or photon location.

Although it is argued in this paper that current teleportation
experiments cannot lead to reliable teleportation procedure and that
for alternative proposals some technological tools are currently not
available, we are optimistic about solution of the problem in a
foreseen future. Much efforts made in this direction because
teleportation and related experiments are building blocks of quantum
cryptography and quantum computation: the two extremely important
fields which  are on the verge of
transformation from gedanken ideas  to reality.

  The research was supported in part by grant 614/95
 of the Basic Research Foundation (administered by the Israel Academy
 of Sciences and Humanities). Part of this work was done during the
 1998 Elsag-Bailey Foundation research meeting on quantum computation.

\end{multicols}


\begin{thebibliography}{99} 


\bibitem{OX}
{\em The Oxford  English Dictionary}, 2nd Ed. (Clarendon
Press, Oxford 1989) XVII p. 730.


\bibitem{BBC}
C. H. Bennett, G. Brassard, C. Crepeau, R. Jozsa, A. Peres, 
W. K. Wootters,  Phys. Rev. Lett.  {\bf 70}, 1895 (1993).

\bibitem{Inn}
B. Bouwmeester, J. W. Pan, K. Mattle, M. Eibl, H. Weinfurter, 
A. Zeilinger,  Nature {\bf 390}, 575 (1997).

\bibitem{Inn-mix}
J. W. Pan, D. Bouwmeester, H. Weinfurter, and A. Zeilinger 
 Phys. Rev. Lett. {\bf 80}, 3891  (1998). 

\bibitem{Rome}
D. Boschi, S. Branca, F. De Martini, L. Hardy and S. Popescu,
  Phys. Rev. Lett. {\bf 80},  1121  (1997).

\bibitem{tele-V}
 L.Vaidman,   Phys. Rev.  {\bf A  49}, 1473 (1994).
 
\bibitem{Norbert} Before completing the manuscript we have learned
  that J. Calsamiglia and N.  L\"utkenhaus (unpublished) considered
  independently this proof for photons. They also proved impossibility
  of the nondegenerate Bell measurement for the case which is not
  considered here. They allowed indirect interaction between quantum
  particles by making measurements in two stages such that the choice
  of the measurements in the second stage depends on the results of
  the measurements in the first stage.

\bibitem{Harold}
 H. Weinfurter,  Europhys. Lett. 559 (1994).

\bibitem{BrMa}
 S. L. Braunstein and A.
Mann,  Phys. Rev. A {\bf 51}, R1727 (1955).

\bibitem{Mich}
M. Michler, K. Mattle, H. Weinfurter, and
    A. Zeilinger, Phys Rev. A {\bf 53}, R1209 (1996).

\bibitem{Zei-trie}
 K. Mattle, M. Eibl, H. Weinfurter, 
A. Zeilinger,  in {\em Quantum Interferometry}, F. De Martini,
G. Denardo  and Y. Shih, eds., (VCH, Weinheim 1996) p. 57.

\bibitem{AAV86}
Y. Aharonov, D. Albert, and L.Vaidman,
 Phys. Rev. D {\bf
34}, 1805 (1986).

\bibitem{Pop}
S. Popescu, e-print quant-ph/9501020.

\bibitem{Slea}
 T. Sleator and H. Weinfurter, in {\it IQEC Technical
    Digest 1994}, Vol. 9, 1994, OSA Technical Digest Series (OSA,
  Washington, D.C., 1994), p. 140. 

 \bibitem{David}
      L. Davidovich, N.  Zagury, M. Brune, J. M. Raimond and S. Haroche,
   Phys. Rev. A,  {\bf 50}, 895, (1994).

\bibitem{Cirac}
 J. I. Cirac and A. S. Parkins,
 Phys. Rev. A {\bf 50} R4441 (1994).

\bibitem{ZG97}
 S. B.  Zheng, G. C.  Guo, 
  Phys. Lett.  A {\bf 232},  171 (1997). 

\bibitem{Har}
S.  Haroche, M. Brune, J. R. Raimond,
 Philos. T. Roy. Soc. A {\bf 355} (1733) 2367 (1997). 

\bibitem{Almei}
N. G. Almeida, L. P. Maia, C. J. Villas-Boas, M. H. Y.  Moussa, 
Phys. Lett.  A {\bf 241},  213 (1998). 

\bibitem{Mous-swa}
M. H. Y. Moussa,
 Phys. Rev. A {\bf 55} R3287 (1997).

\bibitem{BK}
S. Braunstein, H. J. Kimble, Phys. Rev. Lett. {\bf 80}, 869 (1998).

\bibitem{Br-priv}
S. Braunstein, talk at Benasque Center of Physics.

\bibitem{Mol}
S. N. Molotkov,
Phys. Lett.  A {\bf 245},  339 (1998). 


\end{thebibliography}
\end{document}